# Charge-induced energy shift of a single-spin qubit under a magnetic-field gradient

Takashi Kobayashi[1, a)], Akito Noiri[2], Takashi Nakajima[2], Kenta Takeda[2], Leon C. Camenzind[2], Ik Kyeong Jin[2], Giordano Scappucci[3,4], and Seigo Tarucha[1,2]

[1]*RIKEN Center for Quantum Computing, Wako, Saitama 351-0198, Japan*

[2]*RIKEN Center for Emerging Matter Science, Wako, Saitama 351-0198, Japan*

[3]*QuTech, Delft University of Technology, Delft, The Netherlands*

[4]*Kavli Institute of Nanoscience, Delft University of Technology, Delft, The Netherlands*

An electron confined by a semiconductor quantum dot (QD) can be displaced by changes in electron occupations of surrounding QDs owing to the Coulomb interaction. For a single-spin qubit in an inhomogeneous magnetic field, such a displacement of the host electron results in a qubit energy shift which must be handled carefully for high-fidelity operations. Here we spectroscopically investigate the qubit energy shift induced by changes in charge occupations of nearby QDs for a silicon single-spin qubit in a magnetic-field gradient. Between two different charge configurations of an adjacent double QD, a spin qubit shows an energy shift of about 4 MHz, which necessitates strict management of electron positions over a QD array. We confirm a correlation between the qubit frequency and the charge configuration by using a postselection analysis.

## I. INTRODUCTION

Spin qubits based on silicon quantum dots (QDs) are regarded as a compelling platform for quantum information processing. Recently, milestones toward fault-tolerant quantum computation such as fault-tolerant fidelities of universal gates [1–4], measurement and initialization [4–6], and quantum phase-error correction [7] have been achieved. Efforts to scale up toward quantum information processors have spurred extensive research in various areas, including multi-qubit development [6,8], high-temperature operations [4,9–11], and adaption of industrial fabrication technologies [8,12]. In scaling up quantum processors, crosstalk effects from various sources are one of serious obstacles [6]. Spin-qubit operations accompanied by transfer of electrons can cause crosstalk originating from the Coulomb interaction among qubit-host electrons. This class of operations includes qubit shuttling [13–18] and midcircuit qubit measurement for feedback [4,6,19], both of which play indispensable roles in varieties of silicon-spin-based quantum architectures [20–24]. An electron transferred by such operations changes electrostatic potential in nearby QDs due to the Coulomb interaction, displacing electrons inside these QDs. This results in Zeeman energy shift of the spin qubits hosted by the displaced electrons in the presence of inhomogeneous magnetic field [Fig. 1(a)]. In a spin qubit utilizing a magnetic-field gradient induced by a micromagnet [25], this shift leads an excess phase large enough to be a major qubit error source unless treated. To assess the impact of the displacement of a qubit-host electron, the charge-induced energy shift of a spin qubit in a magnetic-field

---

a) takashi.kobayashi@riken.jp

gradient should be proved directly, while a charge-induced change of exchange coupling has been utilized for operations of singlet-triplet qubits before [26–28].

In this work, we report spectral shifts of single-spin qubits induced by displacement of the host electrons inside QDs in the presence of a magnetic-field gradient. The electron position in a QD changes with the electrostatic potential, which is determined by gate biases and charge occupations of surrounding QDs in a QD array. To distinguish and separately investigate the energy shifts induced by changes in charge configurations and gate biases, we implement a protocol to prepare different charge configurations at an identical gate bias set. Measuring qubit resonance frequencies using this protocol, we observe an energy shift of around 4 MHz between two different charge configurations at the same gate bias condition. Measurements with changing gate biases show that this charge-induced energy shift is less sensitive to gate biases than the gate-induced energy shift. Finally we confirm the correlation between the charge configuration and the qubit energy by a postselection analysis.

The device is a triple-QD device made from a $^{28}$Si/SiGe quantum well [Fig. 1(b) left]. The device structure is nominally identical to Refs. [1,17,29]. Voltages on the plunger gates labelled P1, P2, and P3 predominantly change the electric potential of the QDs (QD1, QD2, and QD3) and thus their electron occupations ($n_1$, $n_2$, and $n_3$). The pitch of the plunger gates is 90 nm. The barrier gates B1, B2, B3, and B4 are used to modify tunnel coupling between a QD and an electron reservoir (for B1 and B4) or between QDs (for B2 and B3). The barrier voltages are kept constant throughout the experiments. We denote charge configurations of the triple-QD system as $n_1n_2n_3$. The device is tuned to the three-electron regime as shown in the charge stability diagram in Fig. 1(c), where QD-reservoir and inter-QD degeneracy points (solid and dotted lines) separate charge-stable regions labelled by the ground-state charge configurations (102, 111, and 201). We refer to the spin qubits hosted by electrons in QD$i$ ($i$ = 1, 2, 3) as q$i$, while q2 is not used in this work. The device is subjected to an external magnetic field of 420 mT (white arrow). A cobalt micromagnet fabricated on top of the device [Fig. 1(b) right] induces a longitudinal magnetic-field gradient over the QD array. The total magnetic field increases from left to right, and gives qubit energies of around 16 GHz. The QD array is nominally on the symmetry axis of the micromagnet (dashed line) to suppress a magnetic-field gradient along the vertical axis. A transverse magnetic-field gradient is also induced, enabling us to electrically drive spin resonance by applying microwave voltage to a gate under the plunger and barrier gates. Valley splittings in QD1, QD2, and QD3 measured by magneto-spectroscopy are 80 μeV, 100 μeV, and 90 μeV, respectively, large enough to suppress population of excited valley states. Transitions of charge configurations are manifested as sudden changes in conductance of the adjacent charge sensor QD, which is measured by radio-frequency (rf) reflectometry [30]. A tank circuit for the reflectometry consists of a NbTiN spiral inductor with inductance of 2.1 μH and a parasitic capacitance, and we use the probe-tone frequency of 187.6 MHz. The qubit states are measured by a combination of energy-selective tunneling to a reservoir and the charge-sensing measurement [31]. The device is cooled down by a dilution refrigerator with the base temperature of 30 mK while the electron temperature is 70 mK.

## II. CHARGE-PREPARATION PROTOCOL

To measure the charge-induced qubit energy shift in the identical gate-bias condition, we utilize a pulse sequence to prepare the 111 configuration in the 102 and 201 charge-stable regions as shown in Fig. 1(d) [also see Fig. 1(c) for the actual voltage set used for $V_1$ and $V_2$]. The system is first ramped adiabatically from the 102 (201) region to the 111 region, and then pulsed back to the 102 (201) region. Pulsed back non-adiabatically against the inter-QD tunnel coupling, the system can remain in an excited state having the 111 configuration in the 102 or 201 region regardless of the spin state. The device is tuned to suppress the tunnel couplings between neighboring QDs. This enables us to implement a non-adiabatic passage by using a short pulse-back duration $\tau_{Back}$. We note that the actual ramp rate is limited by low-pass filters with cutoff frequency of 39 MHz. The passage to the 111 region is kept adiabatic by slowly ramping gate biases over a narrow range (10 mV in the P1 or P3 bias) around the inter-QD charge degeneracy point in a long duration of 100 μs.

Figure 1(e) shows two traces of charge-sensor signals $V_{rf}$ as a function of time after returning to the 102 region $t_{Record}$, evaluating the charge-preparation protocol with $\tau_{Back} = 0$. The blue and grey traces are typical results indicating occupation of the 111 and 102 configurations, respectively. Figure 1(f) shows a histogram of $V_{rf}$ at $t_{Record} = 0$ (blue bars). Using a threshold value to distinguish the 111 and 102 configurations (blue dashed line), we estimate a probability to obtain the 111 configuration $P_{111}$ at 84 % in the 102 region. We perform a similar experiment in the 201 region (orange bars) and obtain $P_{111}$ = 68 % due to larger tunnel coupling between QD1 and QD2. The abrupt drop of the signal in the blue trace (red arrow) indicates the relaxation from the 111 configuration to the 102 configuration. We extract the relaxation time of the 111 configuration $T_{1c}$ from the statistics of the jump times shown in Fig. 1(g), finding $T_{1c}$ = 3.3(1) ms in the 102 region and $T_{1c}$ = 10.2(7) ms in the 201 region. The difference in $T_{1c}$ can be explained by difference in the QD-reservoir tunnel couplings for QD1 and QD3. Since these $T_{1c}$ values are much longer than the spin-qubit operation time for spectroscopy, we can measure the resonance frequencies of spin qubits in the 111 configuration even in the 102 and 201 regions before the charge configuration relaxes to the ground state.

## III. QUBIT SPECTROSCOPY WITH CHARGE PREPARATION

Figure 2(a) schematically shows the sequence to manipulate spin qubit q1 with the prepared charge configurations in QD2 and QD3. The q1 state is first initialized to the spin-down state in the 102 region using energy selective tunneling between QD1 and the adjacent reservoir. Then the 111 or 102 configuration is prepared in the 102 region. During a wait time of 30 μs, microwave pulses are applied to manipulate q1. The exchange coupling between q1 and q2 in the 111 configuration is negligibly small in the middle of the 102 region owing to the tunnel coupling suppressed for the charge preparation. Before the energy-selective readout of the q1 state, we reset the charge configuration to 102, since the optimum readout point depends on the charge configuration. For this purpose, bringing the gate biases to $V_3$ in Fig. 1(c), relaxation from the 111 configuration to the 102 configuration is facilitated. This enables us to read out q1 in the 102 configuration near the 102-002 charge degeneracy point [$V_4$ in Fig. 1(c)] independently of the charge configuration during qubit operations. Interchanging roles of QD1 and QD3, we can operate q3 with the charge preparation. We implement this experimental sequence by the pulse sequence shown in Fig. 2(b). A reset-stage duration $\tau_{Reload}$ of 5 us (15 us) is used for operations of q1 (q3).

Microwave response of q1 and q3 exhibits charge-configuration dependence as shown in Figs. 2(c-h). The Rabi chevron pattern for q1 (q3) in Fig. 2(c) [(f)] is taken without bringing the system from 102 (201) to the 111 region as a reference, centered around 15.581 GHz (16.425 GHz). We observe no alias features due to occupation of the excited valley states [32]. Figures 2(d) and (g) show the microwave response of q1 and q3, respectively, when the charge configurations are prepared with $\tau_{Back}$ = 0. The chevron pattern of q1 shifts to a lower frequency (15.577 GHz), and the q3 spectrum shows an alias feature at a higher frequency (16.429 GHz). The exchange interactions from q2 are suppressed by the small inter-QD tunnel couplings, hardly accounting for these results. $\tau_{Back}$ of about 1 μs or longer makes the pulse-back stage adiabatic against the inter-QD tunnel coupling; the Rabi chevron patterns in the reference experiments revert with $\tau_{Back}$ = 1 μs [Figs. 2(e) and (h)]. This indicates that the prepared charge configuration determines the spectra.

The charge-configuration dependence of the qubit resonance frequencies is explained by the Coulomb repulsion in the presence of the magnetic-field gradient. The q1-(q3-)host electron is repulsed leftward (rightward) when an electron moves from QD3 (QD1) in the 102 (201) configuration to QD2 in the 111 configuration. Since the local magnetic field increases from left to right, these repulsions by the 111 configuration should result in a negative frequency shift for q1 and a positive shift for q3 consistently with the observed frequency shifts. We estimate that the displacement of the electron hosting q1 is 0.9 nm using the measured frequency shift of 4 MHz and the average magnetic field gradient of 170 μT/nm. The field gradient is calculated from the difference of resonance frequencies between q1 and q3 [Figs. 2(c) and (f)] and the distance between QD1 and QD3 expected from the plunger-gate pitch. We compare this estimation with a simple electrostatic model. We consider displacement of an electron in a harmonic potential with an orbital level spacing of $h\omega_{orb}/2\pi$ ($h$ is the Plank constant) by the Coulomb potential from another electron fixed at a distance $d$ away. Treating the electrons as point charges, the former electron is displaced by $\Delta x(d) = e^2/4\pi\varepsilon m^*\omega_{orb}^2 d^2$ ($e$ is the elementary charge, $m^*$ is the electron effective mass, and

$\varepsilon$ is dielectric constant in silicon) for $\Delta x(d) \ll d$ in comparison with the absence of the latter electron. The electron displacement between the 102 and 111 configurations is expressed by $\Delta x(d) - \Delta x(2d)$. For a silicon QD with a typical orbital spacing of 1 meV and $d$ = 90 nm, we obtain $\Delta x(d) - \Delta x(2d)$ of about 4 nm. Screening of the Coulomb interaction by gate electrodes may account for the discrepancy between the simple electrostatic model and the experiment.

Qubits defined in different charge configurations can be susceptive to gate biases differently, which can be probed by measuring the resonance frequencies as a function of gate biases. Figure 3(a) schematically shows the pulse sequence to measure the gate-bias dependence of the q1 resonance frequencies, where the gate bias on the qubit-operation stage (colored segment) is varied along the white arrow in Fig. 3(b) so that the energy levels in QD1 and QD2 are detuned without moving the QD2 level. For this experiment, a moderate $\tau_{Back}$ value of 0.2 µs is used to evenly populate the 102 and 111 configurations. In the operation stage, we apply a Ramsey pulse sequence consisting of two $\pi/2$ pulses with a microwave frequency $f_0$ separated by an evolution time $t_{evol}$ to precisely extract q1 resonance frequencies in the 102 and 111 configurations $f_{102}$ and $f_{111}$, respectively. Figure 3(c) displays superimposed Ramsey fringes as a function of the energy level detuning, showing two oscillation modes attributed to the 102 and 111 configurations. Here the detuning value on the perpendicular axis is indicated by the P1-bias offset from $V_1$ [white circle in Fig. 3(b)], $\Delta$P1. By fitting a function $A_1\exp[-(t_{evol}/T_2^*)^2]\cos[2\pi\delta_1 t_{evol}+\phi_1] + A_2\exp[-(t_{evol}/T_2^*)^2]\cos[2\pi\delta_2 t_{evol}+\phi_2] + B$ to the Ramsey fringes at each $\Delta$P1, we obtain the oscillation frequencies $\delta_1$ and $\delta_2$ corresponding to $|f_{102} - f_0|$ or $|f_{111} - f_0|$. Combining datasets measured with $f_0$ = 15.582 GHz [Fig. 3(c)], 15.575 GHz, and 15.572 GHz (not shown), we reconstruct the $\Delta$P1 dependence of $f_{102}$ and $f_{111}$ from $\delta_1$ and $\delta_2$ as shown in Fig. 3(d).

The q1 resonance frequencies depend on the QD level detuning almost linearly in both charge configurations, which is attributed to electron displacement by gate biases. In the detuning range between the 012-102 and 102-111 charge degeneracy points (dotted lines), $f_{102}$ and $f_{111}$ are well fitted by linear functions with slopes of 0.090(2) MHz/mV (blue line) and 0.101(3) MHz/mV (red line). The similar slopes indicate small gate-bias sensitivity of the charge-induced energy shift. A residual detuning dependence appears in a frequency difference $f_{102} - f_{111}$ [Fig. 3(e)]. While the difference almost stays within the range between 4.0 MHz and 4.5 MHz in the 102 region, we obtain a slope of −0.011(2) MHz/mV by a linear fit (orange line). This small detuning dependence can be attributed to quadraticity of the magnetic field induced by the micromagnet [33,34], or change in the displacement of the q1-host electron. We note that the linear dependence of $f_{102} - f_{111}$ over a wide range of $\Delta$P1 implies negligible contribution of exchange interaction between electron spins in QD1 and QD2. We can also measure $f_{102}$ and $f_{111}$ in the 012 and 111 regions owing to the suppressed tunnel couplings. $f_{102}$ and $f_{111}$ change with $\Delta$P1 near the 102-

111 transition sharply, which could be due to the change in charge distribution over QD2 and QD3 induced by the inter-QD tunnel coupling.

We find that the gate biases and the charge configuration contribute to the qubit energy and the single electron level differently. On one hand, over the $\Delta$P1 range of 40 mV, $f_{102}$ and $f_{111}$ change by about 4 MHz. On the other hand, the same $\Delta$P1 range shifts the single-electron energy level of QD1 about 3 meV. The difference between the 102 and 111 configurations, which induces a comparable qubit energy shift (about 4 MHz), shifts the single-electron energy level of QD1 by 0.4 meV estimated from the charge stability diagram. This comparison indicates that the inter-QD Coulomb interaction can shift the q1 resonance frequency strongly even if it shifts the single electron level weakly. This is presumably because the Coulomb interaction moves electron along the direction of the magnetic field gradient more effectively than the plunger gate biases in the present device. Qubit operations accompanied by electron transfers require assessment of the impact of the Coulomb interaction on surrounding qubits, as it could be larger than expected from shift of single electron levels observed in a charge stability diagram.

## IV. CORRELATION BETWEEN CHARGE AND QUBIT FREQUENCY

We confirm the correlation between the resonance frequency and the charge configuration by postselecting single-shot qubit-measurement outcomes according to charge-measurement outcomes. Figure 4(a) schematically shows a pulse sequence to perform the postselection. To read out the charge configuration in addition to the qubit readout, the charge-sensor signal is recorded for 25 μs after the qubit operations, with an interim wait time of 25 μs. We perform Ramsey interferometry of q3 using this sequence. The single-shot charge- and qubit-readout signals, $V_{rf,c}$ and $V_{rf,s}$, show histograms as shown in Figs. 4(b) and blue bars in (c), respectively. The single-shot charge-(qubit-)readout signals are assigned to the 201 and 111 configurations (spin-up and -down states) as shown on top of the figure using a threshold value denoted by the dashed line. Using the charge-readout outcomes, we post-select qubit-readout outcomes to calculate the spin-up probability $P_{up}$ corresponding to the 201 and 111 configurations (orange and green bars, respectively).

The postselection reveals the correlation between the qubit resonance frequency and the charge configuration. Figure 4(d) shows Ramsey fringes obtained from the datasets with and without the postselection. Without the postselection (blue), $P_{up}$ shows superimposed oscillations with slow and fast frequencies. These oscillation modes separately appear in $P_{up}$ with the postselection on the 201 and 111 configurations (orange and green). This isolation is more obvious in the Fourier transform spectra of the Ramsey fringes measured with changing the microwave frequency [Fig. 4(e)]. While two funnels centered at different microwave frequencies appear in the spectrum without the postselection (left panel), they are well isolated in the spectra with the postselections (middle and right panels). As the center of the funnels corresponds to the resonance frequency, these spectra prove the correlation between the resonance frequency and the charge configuration.

## V. CONCLUSIONS

In conclusion, we have demonstrated spectral measurements of silicon spin qubits with preparing the charge configuration. We observe spectral shifts attributed to displacement of qubit-host electrons under a magnetic-field gradient. The resonance-frequency shift induced by different charge configurations are around 4 MHz, which will cause a large phase accumulation in qubit operations. The correlation between the charge configuration during the qubit operation and the resonance frequency is confirmed by a postselection analysis. These results reveal the impact of the charge configuration of QDs surrounding a spin qubit utilizing a magnetic-field gradient. To avoid errors due to the charge-induced energy shift, timing of electron transfers must be managed strictly. In qubit shuttling, the transfer timing jitters if a QD energy level in shuttling path fluctuate in time. This timing jitter will be severe in slow shuttling in Ref. [18]. Spin-to-charge conversion for qubit readout [31,35] also hinders the strict management, since the resulting charge configuration depends on the qubit configuration and thus there should not be previous knowledge. To deal with such operations, a sophisticated operation sequence is required to cancel phase accumulation unconditionally on the charge configuration as proposed in Ref. [5]. The exchange interaction can also cause spin-dependent charge configuration [26–28] if the two-qubit system is near the inter-QD charge degeneracy point. Qubit operations at the charge symmetry point are important to suppress not only decoherence of the two-qubit system but also impact of the spin-dependent charge configuration to other qubits surrounding it.

## ACKNOWLEDGMENTS

This work was supported financially by JST Moonshot R&D Grant Number JPMJMS226B and JSPS KAKENHI grant Nos. 22H01160 and 23H05455. A.N. acknowledges support from JST PRESTO Grant Number JPMJPR23F8. T.N. acknowledges support from JST PRESTO Grant Number JPMJPR2017. L.C.C. acknowledges support from Swiss NSF Mobility Fellowship No. P2BSP2 200127. We thank Kazumasa Makise for fabrication of the NbTiN spiral inductor.

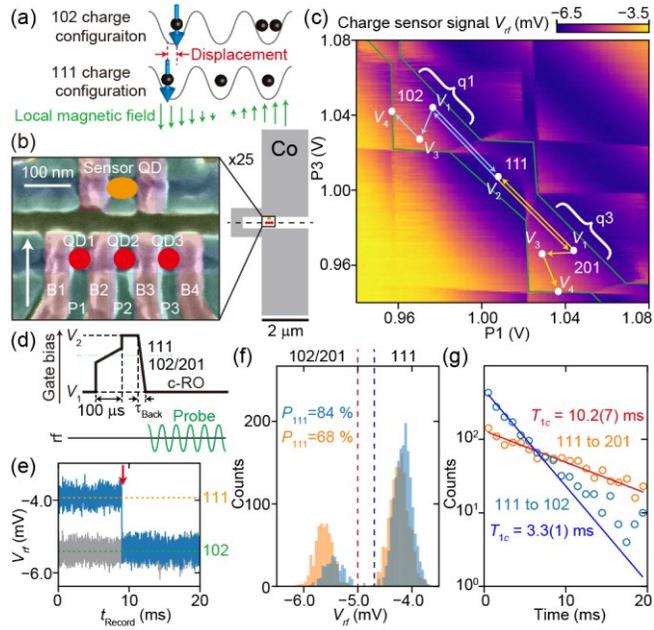

FIG. 1. Charge-preparation protocol. (a) Schematic image of the mechanism of the charge-configuration-dependent energy shift of a spin qubit. (b) SEM image of the device (left) and schematic drawing of the cobalt micromagnet (right). The red circles show QDs hosting spin qubits. The orange ellipsoid shows the QD serving as a charge sensor. The white arrow shows direction of the external magnetic field. The black rectangle in the right panel shows the area shown in the left panel. The micromagnet is inversion symmetric about the dashed line. (c) Charge stability diagram as a function of P1 and P3 biases. Solid and dotted lines show position of charge degeneracy points relevant to QD-reservoir and inter-QD tunneling, respectively. $n_1n_2n_3$ denotes the charge configuration giving the ground state in the region enclosed by the solid and dashed lines. $V_1$, $V_2$, $V_3$, and $V_4$ denotes gate-bias sets used in experiments. For $V_1$, $V_3$, and $V_4$, different bias sets are used between experiments containing q1 and q3 operations as denoted in the figure. The bias set for $V_2$ is same for both q1 and q3 experiments. (d) Sequence of gate-bias (top) and rf pulses (bottom) to evaluate the charge-preparation protocol. The system prepared in the 102/201 region is adiabatically transferred to the 111 region through the charge degeneracy point between the 102/201 and 111 regions (green dotted line), and returned from center of the 111 region ($V_2$) to center of the 102/201 region ($V_1$) in the pulse-back duration $\tau_{\text{Back}}$. To evaluate the probability to obtain the 111-charge configuration $P_{111}$ in the 102/201 region, we apply a probe-tone burst (green meandering curve) and measure reflection in the charge-readout (c-RO) stage. (e) Two typical charge-sensor signals $V_{\text{rf}}$ as a function of time after returning to the 102 region, $t_{\text{Record}}$. The high (low) $V_{\text{rf}}$ value denoted by the orange (green) dashed line indicates occupations of the 111 (102) configuration. (f) Histogram of the $V_{\text{rf}}$ value at $t_{\text{Record}} = 0$. Blue (orange) data are measured in the 102 (201) region. The dashed lines show threshold values used to discriminate charge configurations. (g) Distributions of the dwell time in the 111 configuration in the 102 and 201 regions

(blue and orange). The blue and red curves are exponential fit functions to evaluate the charge relaxation time $T_{1c}$ shown in the figure.

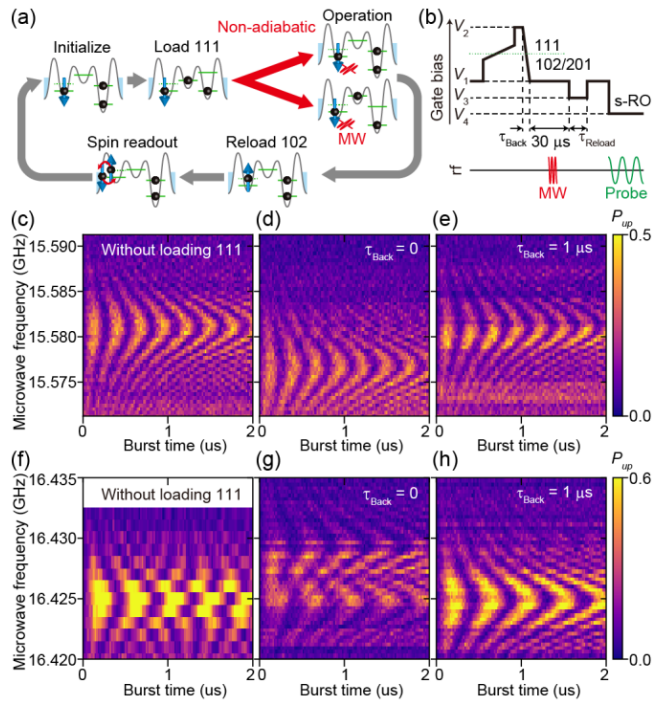

FIG. 2. Qubit spectroscopy with the charge-preparation protocol. (a) Schematic of the q1 operation sequence incorporated with the charge-preparation protocol. (b) Pulse sequence to implement the conceptual sequence in (a). The voltage sets $V_1$, $V_2$, $V_3$, and $V_4$ in the gate-bias pulse sequence are shown in Fig. 1(c). The red meandering curve in the rf pulse sequence shows microwave for qubit control. The qubit state is measured in the spin-readout (s-RO) stage at the end. (c)-(h) Spectroscopy measurements of q1 (c-e) and q3 (f-h) using a microwave burst without the charge preparation as a reference [(c) and (f)], with the charge preparation using $\tau_{Back} = 0$ [(d) and (g)] and $\tau_{Back} = 1$ µs [(e) and (h)].

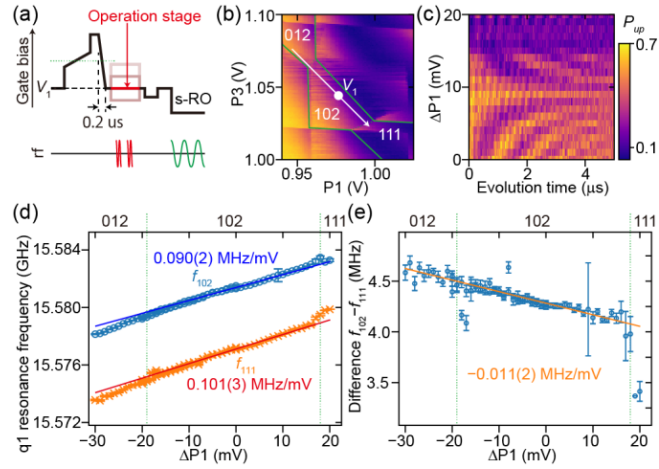

FIG. 3. Detuning dependence of q1 resonance frequencies. (a) Pulse sequence to measure the detuning dependence. Voltage in the operation stage (colored segment) is changed in the experiment. During this segment, two microwave pulses for the Ramsey interferometry are applied to q1. We used $\tau_{Back}$ of 0.2 μs. (b) A magnification of the charge stability diagram in Fig. 1(c) around the 102 region. The white arrow shows the axis of energy-level detuning between QD1 and QD2, and the white circle shows its origin corresponding to $V_1$ in Fig. 1(c). The green dotted lines show the 012-102 and 102-111 charge degeneracy points, respectively. (d) Spin-up probability $P_{up}$ obtained from the Ramsey interferometry as a function of detuning. Detuning is indicated by variation of the P1 gate bias from $V_1$, $\Delta P1$. The microwave frequency is fixed at 15.582 GHz. The resonance frequency in the 102 (111) configuration $f_{102}$ ($f_{111}$) is plotted as the blue (orange) symbols. The linear fit curves to $f_{102}$ and $f_{111}$ are shown by the blue and red solid lines, whose slopes are denoted by the same-colored numbers. The detuning range is separated to the 012, 102, and 111 regions as denoted above the top axis. The left and right dotted lines show the $\Delta P1$ values at the 012-102 and 102-111 charge degeneracy points. (e) Detuning dependence of the frequency difference $f_{102} - f_{111}$ (blue symbols). The solid line shows the linear fit curve to $f_{102} - f_{111}$, having a slope of −0.011(2) MHz/mV as denoted in the figure.

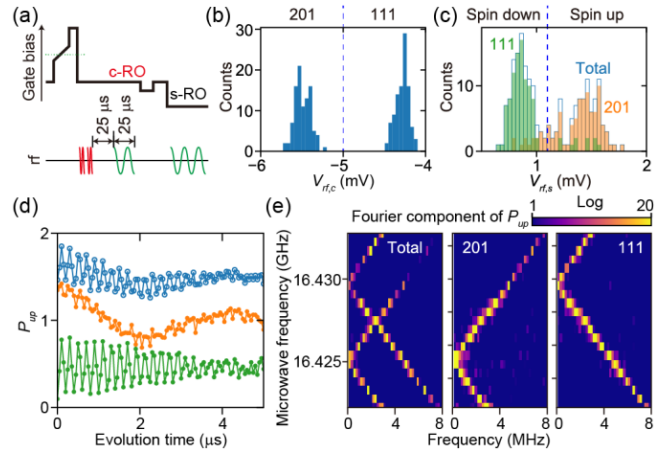

FIG. 4. Ramsey interferometry of q3 with postselection according to the charge configuration. (a) Pulse sequence for the experiment. A probe-tone pulse for an additional charge-readout stage (c-RO in red) is inserted between the second microwave pulse and the charge reset stage. (b), (c) Histograms of the single-shot charge- and qubit-readout signals, $V_{rf,c}$ [(b)] and $V_{rf,s}$ [(c)]. $V_{rf,c}$ is an averaged value of the charge-sensor signal $V_{rf}$ for 25 μs, and $V_{rf,s}$ is a peak-to-peak value of $V_{rf}$ over the qubit-readout duration. The dashed lines in (b) and (c) shows threshold values used to assign the charge- and qubit-readout signals to the outcomes denoted above the top axes. The blue bars in (c) show the total qubit-readout signals. Orange and green bars are subsets of the total signal postselected on the charge-readout outcome of the 201 and 111 configuration, respectively. (d) Typical Ramsey fringes without postselection (blue) and with postselection on the 201 and 111 configurations (orange and green). The microwave frequency is 16.425 GHz. (e) Fourier transformed Ramsey fringes without postselection (left panel) and with postselection on the 201 and 111 configurations (center and right panels).